\documentclass[conference]{IEEEtran}
\IEEEoverridecommandlockouts

\usepackage{cite}
\usepackage{amsmath,amssymb,amsfonts}
\usepackage{algorithmic}
\usepackage{graphicx}
\usepackage{float}
\usepackage{booktabs}
\usepackage{soul}
\usepackage{textcomp}
\usepackage{xcolor}
\def\BibTeX{{\rm B\kern-.05em{\sc i\kern-.025em b}\kern-.08em
    T\kern-.1667em\lower.7ex\hbox{E}\kern-.125emX}}
\begin{document}

\title{Toward Temporal Realism in City-Scale Crisis Response Simulation using LLM Agents}

\author{
\IEEEauthorblockN{
Anping Zhang\textsuperscript{1,2},
Yang Tan\textsuperscript{1},
Yuanbo Tang\textsuperscript{1},
Huaze Tang\textsuperscript{1},
Qiuhua Ye\textsuperscript{1},
Marta C. Gonzalez\textsuperscript{2},
Yang Li\textsuperscript{1,3}\thanks{Corresponding author: yangl@cuhk.edu.cn}
}
\IEEEauthorblockA{
\textsuperscript{1} Tsinghua Shenzhen International Graduate School, Tsinghua University\\
\textsuperscript{2} Department of City and Regional Planning, University of California, Berkeley\\
\textsuperscript{3} School of Artificial Intelligence, The Chinese University of Hong Kong
}
}
\maketitle

\begin{abstract}
Human collective participation is rarely steady in time: it is bursty, with short episodes of intense activity separated by long quiet intervals. In crisis response and community mobilization, predicting \emph{when} people act matters as much as predicting \emph{whether} they act. Such settings are increasingly modeled with large language model (LLM) based social simulators, yet these simulators are validated on whether each action is individually plausible, not on whether actions are timed as in reality. Their \emph{temporal realism}, the degree to which simulated activity reproduces the bursty, heavy-tailed timing of real human systems, thus remains untested. We examine this gap using a multi-year, city-scale log of offline volunteering in Shenzhen that spans the COVID-19 pandemic. Empirically, we establish that bursty timing is common at both the individual and the tracked-group level, that it is largely endogenous and self-exciting, and that it is amplified by the pandemic rather than produced by daily activity cycles. A standard LLM-only simulator reproduces almost none of this timing: its synchronous schedule has no self-excitation channel, so agents act on a near-regular clock. Guided by these findings, we build a simulator in which a data-calibrated self-excitation channel and a crisis-period regime decide \emph{when} each agent acts and query the LLM only at those moments, leaving it to decide \emph{which} task to join and \emph{whether} to commit. The LLM-only baseline yields no bursty agents (median burstiness $B=-0.14$); as a proof-of-concept, a single data-calibrated gate is then sufficient to lift per-agent timing above the burst threshold (median $B\approx0.37$) without degrading the LLM's content decisions. These results indicate that temporal realism in LLM-based crisis-response simulation is best achieved by decoupling \emph{when} agents act, governed by an explicit self-excitation and crisis-activation mechanism, from \emph{what} they do, governed by the LLM.
\end{abstract}


\section{Introduction}
Human activities often exhibit bursty timing, in which short periods of intense activity are separated by long intervals of inactivity, producing heavy-tailed inter-event time distributions rather than the independent, exponentially distributed waiting times of a Poisson process \cite{harder2006correlated,barabasi2005origin}. While burstiness is well analyzed for individual behaviors and small-group interactions, we still know little about how it appears in large-scale offline collective participation, including volunteering for public events, disaster relief, and community services. This gap matters because the timing of such participation is operationally consequential: whether volunteers arrive in synchronized surges or in a steady trickle determines staffing adequacy during emergencies and how quickly community services can absorb sudden demand.

Recent progress in large language models has enabled LLM-based multi-agent simulation, an increasingly common low-cost testbed for studying group-level social phenomena when field experiments are slow or costly~\cite{gao2024large,park2023generative}. Most such simulators, however, are evaluated on \emph{semantic plausibility}, \emph{individual decision quality}, or \emph{aggregate outcomes}, while \emph{temporal realism}, the degree to which simulated activity reproduces the heavy-tailed and clustered timing observed in real human systems, has received much less attention~\cite{piao2025agentsociety,mou2024unveiling}. This omission is consequential for the very use cases these simulators target: an agent population that acts on a smooth, near-Poisson schedule will systematically miss the abrupt mobilization surges that define real crises, weakening any policy conclusion drawn from such testbeds.

This paper takes a data-driven path from measurement to mechanism to a mechanism-augmented simulator. We organize the work around two goals. First, \emph{measure} whether and how burstiness manifests in large-scale offline volunteering, and isolate which temporal factors drive it. Second, \emph{test} whether injecting the mined mechanism into an LLM-based city simulator is sufficient to lift its timing above the burst threshold, benchmarking against a standard synchronous LLM simulator that serves to confirm the expected timing deficit. Two background gaps motivate this agenda. \textbf{(G1) Empirical gap.} Empirical studies of social bursts have mainly focused on online activities or sensor-based small-group interactions, leaving open how burstiness scales to large offline volunteer collectives and how it is shaped by real-world temporal factors such as circadian rhythms and pandemic-period crisis shocks. \textbf{(G2) Methodological gap.} Many LLM-based multi-agent simulators adopt synchronous or turn-based decision schedules and lack explicit temporal mechanisms such as self-excitation or crisis-driven activation. Absent such a channel, a synchronous schedule reduces each agent to near-independent per-step decisions and is therefore expected to yield near-Poisson rather than heavy-tailed, clustered timing, regardless of prompt quality~\cite{mou2024unveiling,piao2025agentsociety}. The open challenge is therefore not \emph{whether} a vanilla LLM simulator is bursty but \emph{how} to inject the missing mechanism while preserving the LLM's decision quality.

\begin{figure}[h]
\centering
\includegraphics[width=0.85\linewidth]{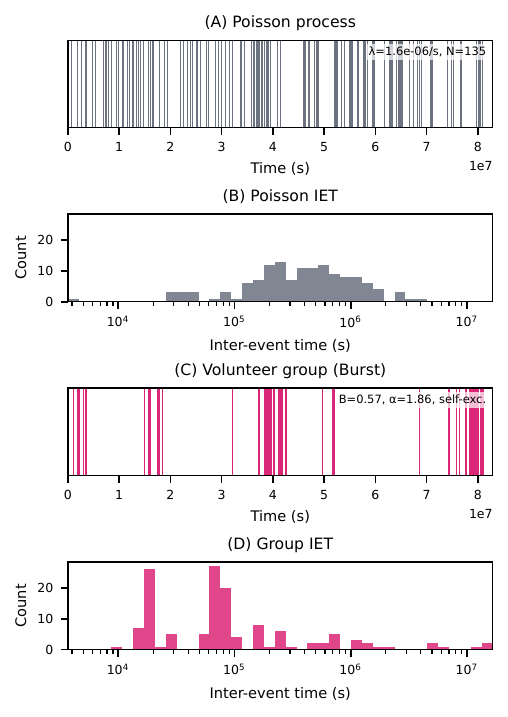}
\caption{An example of bursty participation in a volunteer group. Comparison of inter-event times between a Poisson and a non-Poisson process; all time axes are in seconds. (A) A generated homogeneous Poisson process with rate $\lambda = 1.6 \times 10^{-6}\,\text{s}^{-1}$, matched to the same event count $N=135$ and observation horizon as the real group. (B) Inter-event times of the generated Poisson process. (C) The activity trajectory of a real volunteer group over $\sim$3 years (a representative Burst-classified group with burstiness $B = 0.57$, power-law exponent $\alpha = 1.86$, and a self-exciting Hawkes fit). (D) Inter-event times of the volunteer group. Panels B and D share the same horizontal and vertical scales. The Poisson process shows even spacing concentrated around a single scale, whereas the volunteer group shows clustered timing with many short inter-event times and a few much longer gaps spanning several orders of magnitude.}
\label{fig:burst_poisson_and_non_poisson}
\end{figure}

We ground the study in a Shenzhen volunteering dataset of 6.6 million participation records from $361{,}114$ volunteers spanning 2020--2023 \cite{zhang2022optimising,zhang2026evolution,chen2023volunteer}. We characterize bursty timing at both the individual and the dynamically tracked group level, examine how it shifts under circadian and pandemic-period effects, and attribute it to its generating mechanism. Figure~\ref{fig:burst_poisson_and_non_poisson} contrasts Poisson timing with a real volunteer group trajectory: the Poisson example (Figure~\ref{fig:burst_poisson_and_non_poisson}A--B) shows events spread over time with inter-event times concentrated around a stable scale, whereas the volunteer group (Figure~\ref{fig:burst_poisson_and_non_poisson}C--D) exhibits long inactive gaps interspersed with short clusters of repeated participation, leading to inter-event times spanning several orders of magnitude under the same vertical scale. This contrast motivates treating volunteer participation as a non-Poisson, heavy-tailed process.

Three findings emerge from the measurement. \textbf{(F1)} Burstiness is prevalent: 66.8\% of active individuals and 40.7\% of tracked group trajectories satisfy a joint burst criterion at the task-event level, with heavy-tailed inter-event-time distributions at both levels. \textbf{(F2)} Of two candidate temporal modulators, restricting to daytime active hours leaves the burstiness distribution essentially unchanged, whereas the COVID-19 pandemic period substantially elevates it at the individual level. \textbf{(F3)} The bursts are largely endogenous: about 79\% of bursty volunteers are self-exciting, and the data sit in the near-memoryless region of the burstiness--memory plane, the signature of self-exciting cascades. Guided by these findings, we evaluate an LLM-based city simulator on the AgentSociety \cite{piao2025agentsociety} platform and augment it with a Hawkes-inspired self-excitation channel and a crisis-period regime derived from the pandemic window, gating the LLM decisions with the empirically-estimated self-excitation structure and calibrating its event volume to the simulated horizon.

The contributions of this paper are:
\begin{itemize}
    \item \textbf{A large-scale empirical characterization of offline collective burstiness.} Using 6.6M events from $361{,}114$ Shenzhen volunteers, we show that heavy-tailed, clustered timing is prevalent at both the individual and the dynamically-tracked group level, is crisis-amplified rather than a by-product of circadian cycles, and is predominantly driven by endogenous self-excitation (F1--F3).
    \item \textbf{A \emph{when}/\emph{what}-decoupled simulator for temporal realism.} \emph{Diagnosis.} A synchronous LLM-only simulator structurally lacks any self-excitation channel and therefore stays near-regular regardless of compute (F4). \emph{Feasibility.} As a proof-of-concept, a single data-calibrated timing gate is sufficient to lift the per-agent timing of an offline, city-scale volunteering simulator above the burst threshold without degrading the LLM's content decisions (F5). \emph{Architecture.} The resulting dual-channel protocol decouples \emph{when} an agent acts, set by the Hawkes gate, from \emph{what} it does and \emph{whether} it commits, set by the LLM; a gate-only ablation shows the two are orthogonal and separable (F6). 
\end{itemize}

\section{Related Work}
\label{sec:related-work}

\subsection{Bursty Human Dynamics}
\label{subsec:rw-burst}

Bursty timing has been studied extensively across diverse human behaviors. Foundational work characterizes inter-event-time distributions and introduces the burstiness index $B = (\sigma - \bar{\Delta t})/(\sigma + \bar{\Delta t})$ as a scale-free measure that distinguishes Poisson-like timing from heavy-tailed bursts \cite{goh2008burstiness,karsai2018bursty,barabasi2005origin}. Online communication has been a primary testbed: email, mobile call, and short-message logs robustly exhibit power-law inter-event-time tails, attributed to priority queueing, bounded rationality, or alternating active--inactive states \cite{barabasi2005origin,vazquez2006modeling,oliveira2005darwin,malmgren2008poissonian,gonzalez2008understanding}. Similar heavy tails appear in financial transactions, web browsing, and library access patterns \cite{dezso2006dynamics,bacry2015hawkes}. At the small-group level, bursty patterns also appear in face-to-face contact networks measured by proximity sensors, in temporal interaction networks of academic conferences, and in online discussion threads, where short bursts of repeated interaction alternate with long quiescent intervals \cite{cattuto2010dynamics,isella2011s}.

Mechanism-level explanations of these bursts fall into three complementary families. The first attributes burstiness to \emph{circadian and weekly cycles}, treating heavy tails as a superposition of nearly-Poisson processes operating only during active periods \cite{malmgren2008poissonian,jo2012circadian}. The second attributes burstiness to \emph{endogenous self-excitation}, modeled by Hawkes processes \cite{hawkes1971spectra} whose intensity transiently rises after each event and decays exponentially or as a power law; this family has been used to model retweet cascades, financial trading, and earthquake-like contagion \cite{zhao2015seismic,bacry2015hawkes,ross2021bayesian}. The third treats \emph{exogenous shocks} (news events, policy changes, holidays) as multiplicative perturbations on the baseline rate \cite{crane2008robust,lehmann2012dynamical}.

Despite this breadth, two regimes remain underexplored. First, almost all empirical work targets either online activity streams or sensor-based small-group interactions, leaving large-scale \emph{offline collective participation}, where each event is mediated by a logistical task with capacity, location, and a time window, largely uncharacterized. Second, when circadian and crisis effects can in principle coexist (as during the COVID-19 pandemic), their relative contribution to observed group-level burstiness has not been quantified. Our work addresses both regimes: we characterize burstiness on a 6.6M-event offline volunteering log at the task-event level, and we quantify the relative contribution of circadian and crisis effects, attributing the observed bursts to the underlying mechanism families.

\subsection{LLM-Based Social Simulation}
\label{subsec:rw-llm-mas}

Large language models have recently been combined with multi-agent simulation to model human social behavior at scale~\cite{wang2024survey,xi2025rise,gao2024large}. Generative-agent prototypes equip each agent with persona, memory, planning, and reflection modules in sandbox worlds~\cite{park2023generative,zhang2023exploring,wei2022chain,sun2023adaplanner}, and subsequent work scales this paradigm to city-level environments with real maps and mobility~\cite{piao2025agentsociety,xu2023urban} to study cooperation, norm formation, mobility, and collective decision-making~\cite{li2023camel,chen2023agentverse,wang2023unleashing,gao2023s3,zhou2023sotopia}. The same paradigm has been extended to cyber-physical task environments and to agent-based economic, opinion-dynamics, and epidemiological simulators~\cite{qian2024chatdev,hong2023metagpt,zhang2023building,wang2023voyager,zhu2023ghost,li2023large,williams2023epidemic,suzuki2024evolutionary,de2023emergent}.

Existing evaluations of LLM-based social simulators emphasize three dimensions: \emph{semantic plausibility} (whether agent dialogue and decisions sound human), \emph{individual decision quality} (whether agents follow normative or rational baselines on canonical tasks), and \emph{aggregate alignment} (whether macro-level statistics, such as marginal participation rates or opinion distributions, match observed values)\cite{wang2025user,chen2024persona}. \emph{Temporal realism}, the question of whether the timing of simulated actions reproduces the heavy-tailed, clustered patterns of real human dynamics, has received much less attention. This is in part because the underlying simulators typically schedule decisions on a synchronous turn-based or round-robin clock that, absent an explicit self-excitation channel, tends to produce near-Poisson, non-bursty inter-event-time distributions for individual agents\cite{mou2024unveiling,piao2025agentsociety}.

A few threads add explicit temporal mechanisms to such simulators. Earlier agent-based models already use self-exciting processes, but they are not LLM-driven and calibrate one global trace rather than heterogeneous per-user parameters~\cite{zhao2015seismic,rizoiu2017hawkes}. Prompt-level shock injection perturbs LLM behavior, but does not separate the shock from agents' own dynamics or evaluate inter-event times~\cite{gao2023s3,mou2024unveiling}. Closest to us, concurrent work couples self-exciting processes with LLM agents, yet on online domains and against different targets: one line fits per-user Hawkes activation on \emph{online email networks} and validates circadian periodicity and node-level burstiness~\cite{miao2026can}; another drives an LLM opinion simulator with a Hawkes engine and validates macroscopic opinion trajectories rather than individual inter-event timing~\cite{zhang2026evolution}. We instead study large-scale \emph{offline} collective participation, where each event carries logistical capacity, location, and time-window constraints. In this domain, circadian rhythm is negligible for burstiness while the COVID-19 crisis dominates, so we augment only the self-excitation and crisis channels. Concretely, we fit per-user Hawkes parameters on a 6.6M-event population and inject them as a timing \emph{gate} on the LLM channel, with a separate crisis-period regime calibrated to the per-agent burstiness-index distribution.

\section{Data and Burstiness Quantification}
\label{sec:data-and-quant}

This section introduces the dataset, the event representation shared by all analyses, and the statistics we use to decide whether an entity is bursty and whether its bursts are endogenous. 

\subsection{Dataset and Event Representation}
\label{subsec:data}

We use a volunteer participation log from a Chinese volunteering platform that spans 2020-02-14 to 2023-09-30 \cite{zhang2022optimising,zhang2026evolution,chen2023volunteer}. The dataset contains $N = 6.6$~million events generated by $361{,}114$ unique volunteers in Shenzhen; each event is annotated by a task identifier with start/end time, geographic coordinates, type, and capacity. We represent the activity as a time-stamped event log
\begin{equation}
\mathcal{E} = \{(u_i, q_i, t_i)\}_{i=1}^{N},
\end{equation}
where $u_i \in \mathcal{U}$ is a volunteer, $q_i \in \mathcal{Q}$ is a task, and $t_i$ is the participation start time of event $i$. The observation period is partitioned into $M$ consecutive daily snapshots $W_1, W_2, \dots, W_M$, with $\mathcal{E}_m = \{(u, q, t) \in \mathcal{E} \mid t \in W_m\}$; these daily snapshots are used only for the group-tracking step below. For all burstiness and Hawkes analyses we represent each entity (a volunteer or a tracked group) at the \emph{task-event} level: its activity is the sorted sequence of distinct task participation start times at one-second resolution, so that one task participation contributes exactly one event. Inter-event times are then start-to-start gaps measured in seconds, which avoids the event-count inflation and the resolution ceiling of an hourly grid.

For pandemic-related comparisons, we partition the observation horizon at China's relaxation of dynamic zero-COVID restrictions on $d=\text{2022-12-08}$~\cite{china_covid_10measures_2022}, defining a pandemic window $P=\{t<d\}$ and a post-pandemic window $Q=\{t\ge d\}$. This same boundary $d$ is used for the empirical pandemic-effect analysis (Section~\ref{sec:empirical_findings}) and the per-user Hawkes window fits (Section~\ref{subsec:hawkes-fit}).

\begin{figure}[h]
\centering
\includegraphics[width=0.85\linewidth]{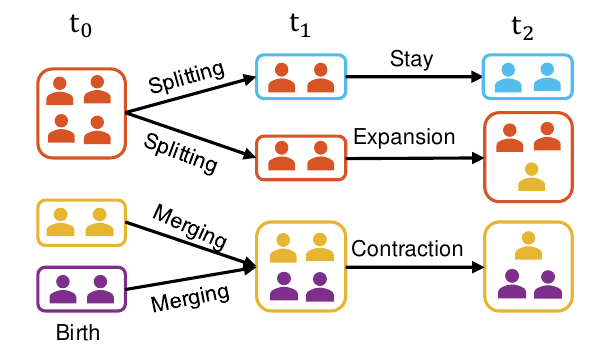}
\caption{Schematic of dynamic-community evolution under daily Jaccard tracking (Table~\ref{tab:dynamic-status}). At each snapshot $t$, a static co-participation group $G^{(t)}$ is matched to existing dynamic communities $F_c$; the lineage records seven event types: \emph{birth} (no match), \emph{stay} (one-to-one match with stable size), \emph{grow/expansion} and \emph{contraction} (one-to-one match with size change), \emph{merge} (many-to-one), \emph{split} (one-to-many), and \emph{death} (prolonged inactivity).}
\label{fig:dynamic_group_tracking}
\end{figure}

\paragraph{Volunteer group construction.}
\label{subsec:group}
We track volunteer groups across snapshots using the daily-Jaccard scheme of Greene et al.~\cite{greene2010tracking}. At each daily snapshot $W_m$, the static group induced by task $\tau$ is the set of co-participants $G_{m,\tau} = \{u \in \mathcal{U} \mid (u, \tau, t) \in \mathcal{E},\, t \in W_m\}$, retaining only $|G_{m,\tau}| \ge 2$. Each new group $G$ is matched against the latest membership set $F_c$ of every active dynamic community $c$ via the Jaccard coefficient
\begin{equation}
J(F_c, G) = \frac{|F_c \cap G|}{|F_c \cup G|},
\end{equation}
and a match requires $J(F_c, G) > \theta$ with $\theta = 0.3$. Birth, stay, grow, contraction, merge, split, and death events are recorded according to Table~\ref{tab:dynamic-status} to maintain the dynamic community lineage; Figure~\ref{fig:dynamic_group_tracking} illustrates the main evolution types.

\begin{table}[H]
\centering
\small
\setlength{\tabcolsep}{8pt}
\caption{Dynamic-community state transitions under Jaccard tracking. Here $r = |G^{(t)}|/|F_c|$ is the size ratio and $|\mathrm{Match}(G^{(t)})|$ is the number of matched communities at snapshot $t$.}
\label{tab:dynamic-status}
\begin{tabular}{lll}
\toprule
\textbf{Status} & \textbf{Condition} & \textbf{Label} \\
\midrule
Birth       & $|\mathrm{Match}(G^{(t)})| = 0$ & new \\
Stay        & one-to-one, $0.9 \le r \le 1.1$ & stay \\
Grow        & one-to-one, $r > 1.1$ & grow \\
Contraction & one-to-one, $r < 0.9$ & contraction \\
Merge       & many-to-one & merge \\
Split       & one-to-many & split \\
Death       & inactive for $\ge s$ snapshots & death \\
\bottomrule
\end{tabular}
\end{table}

\subsection{Quantifying Burstiness and Endogeneity}
\label{subsec:burst-quantification}

\paragraph{Inter-event times.}
For each volunteer $u$, sorting participation start times $t^u_{(1)} < \dots < t^u_{(n_u)}$ (in seconds) yields $\Delta t^u_k = t^u_{(k+1)} - t^u_{(k)}$. For each tracked dynamic group $g$, we collect the start times of all distinct tasks in which its members co-participated, sort them as $\tilde{t}^{\,g}_{(1)} < \tilde{t}^{\,g}_{(2)} < \cdots$ (one task = one event, in seconds), and form inter-event times $\Delta\tilde{t}^{\,g}_j = \tilde{t}^{\,g}_{(j+1)} - \tilde{t}^{\,g}_{(j)}$.

\paragraph{Burstiness index.}
For an inter-event sequence $\{\Delta t\}$ with mean $\bar{\Delta t}$ and standard deviation $\sigma$, the Goh--Barab\'{a}si burstiness index~\cite{goh2008burstiness} is
\begin{equation}
B = \frac{\sigma - \bar{\Delta t}}{\sigma + \bar{\Delta t}} \in [-1, 1].
\end{equation}

\paragraph{Heavy-tail evidence.}
Following Clauset et al.~\cite{clauset2009power} we fit a discrete power law to $\{\Delta t\}$ and compare it to an exponential null using a likelihood-ratio test (LRT), reporting the standardized log-ratio $R$ and its $p$-value $p_\text{plaw}$.

\paragraph{Burst classification rule.}
An entity (a volunteer or a tracked group) is labeled \emph{bursty} if all of the following hold:
\begin{equation}
n_\text{events} \ge 50,\quad B \ge 0.3,\quad R > 0,\quad p_\text{plaw} < 0.05.
\label{eq:burst-rule}
\end{equation}
The threshold of 50 events ensures the tail-fit reliability discussed in~\cite{clauset2009power}.

\paragraph{Endogeneity: Hawkes self-excitation test.}
To decide whether the heavy-tailed timing is driven by endogenous self-excitation rather than mere Poisson clustering, we fit a univariate Hawkes process with exponential kernel
\begin{equation}
\lambda_u(t) = \mu_u + \sum_{t^u_i < t} \alpha_u \exp\!\bigl(-\beta_u (t - t^u_i)\bigr)
\label{eq:hawkes_individual}
\end{equation}
to each entity's task-event sequence and compare it to a homogeneous Poisson null with rate $n/T$ via the LRT statistic $\Lambda = 2(\ell_\text{Hawkes} - \ell_\text{Poisson})$. Because $\alpha = 0$ lies on the boundary of the parameter space, the asymptotic null distribution is the mixture $\tfrac{1}{2}\chi^2_0 + \tfrac{1}{2}\chi^2_1$ and the boundary-corrected $p$-value is
\begin{equation}
p_\text{LRT} = \tfrac{1}{2}\,\mathrm{erfc}\!\bigl(\sqrt{\Lambda/2}\bigr).
\end{equation}
An entity is labeled \emph{self-exciting} if $p_\text{LRT} < 0.05$ and $\hat\alpha > 0$. As a fit diagnostic we use the time-rescaling theorem: under a correctly specified intensity the rescaled durations $z_i = \int_{t_{i-1}}^{t_i} \lambda(s)\,ds$ are i.i.d.\ $\mathrm{Exp}(1)$, which we check with a one-sample Kolmogorov--Smirnov (KS) test (Smirnov asymptotic $p$-value $p_\text{KS}$).

\section{Burstiness in Real-World Volunteer Participation}
\label{sec:empirical_findings}

We now apply the statistics of Section~\ref{subsec:burst-quantification} to the full log. Three findings (F1--F3) characterize the prevalence, the temporal drivers, and the generating mechanism of the observed burstiness; they jointly motivate the simulator design in Section~\ref{sec:methods}.

\subsection{Prevalence of Bursty Participation (F1)}

We classify each active volunteer (resp. tracked group) as Burst or NonBurst based on the joint criterion $B > 0.3$ and a one-sided likelihood-ratio test of power-law against exponential at $p < 0.05$, restricted to entities with at least 50 task-level events over the 2020--2023 window (Section~\ref{subsec:burst-quantification}). Table~\ref{tab:burst_summar_individual_and_group} reports the resulting label distribution. Among entities with enough events to be classified, bursty timing dominates at the individual level (66.8\% of $24{,}234$ active volunteers) and characterizes a substantial share of tracked group trajectories (40.7\% of $3{,}377$ groups).

\begin{table}[H]
\centering
\caption{Burst vs. NonBurst label distribution at the individual and group level under the canonical thresholds ($B > 0.3$, LR $p < 0.05$, minimum 50 task-level events; sec-level start-to-start inter-event times). Robustness across alternative thresholds is reported in the text.}
\label{tab:burst_summar_individual_and_group}
\begin{tabular}{lrrrr}
\toprule
\textbf{Level} & \textbf{\# Burst} & \textbf{\# NonBurst} & \textbf{Total} & \textbf{Burst (\%)} \\
\midrule
Individual & 16{,}185 & 8{,}049 & 24{,}234 & 66.8 \\
Group      & 1{,}374  & 2{,}003 & 3{,}377  & 40.7 \\
\bottomrule
\end{tabular}
\end{table}

\begin{figure}[h]
    \centering
    \includegraphics[width=1.0\linewidth]{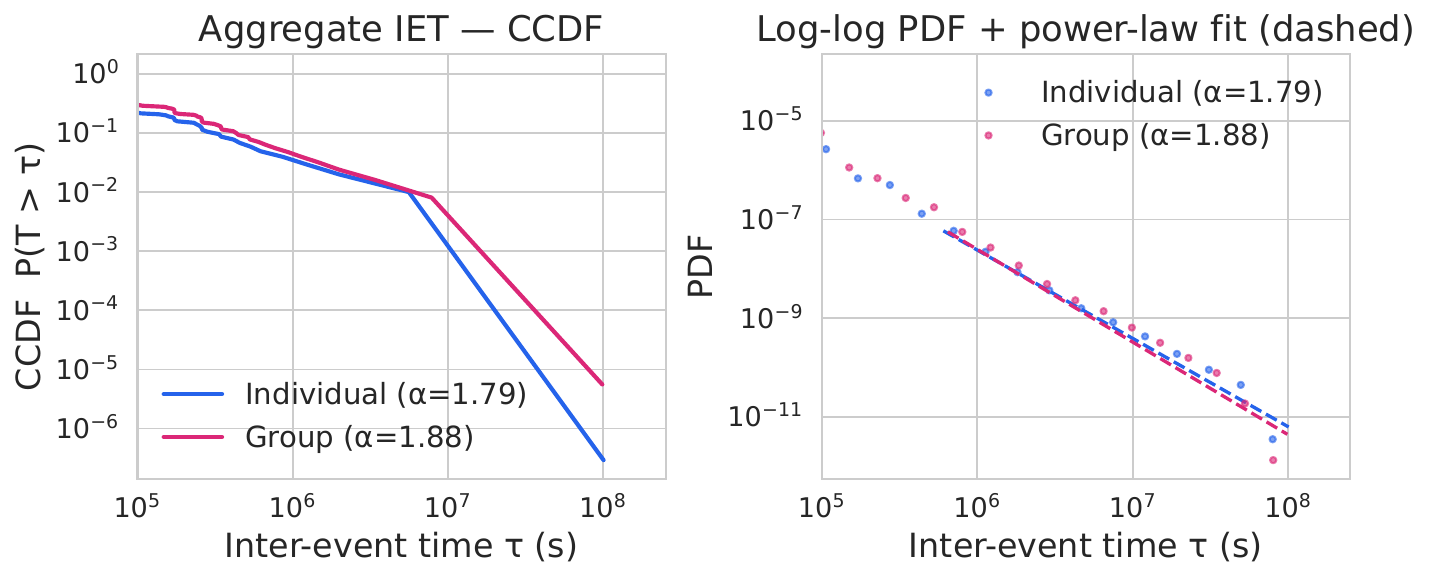}
    \caption{Aggregate inter-event-time (IET) distributions (seconds) for the Burst-classified individuals (aggregate sample $\approx 3.4 \times 10^{6}$ $\tau$ values) and groups ($\approx 1.8 \times 10^{5}$). \textbf{Left:} complementary CDF on log--log axes. \textbf{Right:} empirical PDF on geometrically spaced bins with maximum-likelihood power-law fits (dashed), exponents $\alpha = 1.79$ ($x_{\min} = 6.1 \times 10^{5}\,\text{s} \approx 7\,\text{d}$, individuals) and $\alpha = 1.88$ ($x_{\min} = 6.5 \times 10^{5}\,\text{s} \approx 7.5\,\text{d}$, groups). The display is restricted to the heavy tail $\tau > 10^{5}\,\text{s}$ and the dashed fit line is drawn only over the fitted region $\tau \ge x_{\min}$; the smaller-$\tau$ regime (within-task spillover and the diurnal bump) is not power-law and is omitted. Both power-law fits are decisively preferred over an exponential alternative under a Vuong likelihood-ratio test (normalized log-ratio $R = +44.2$ for individuals and $R = +36.4$ for groups; both $p < 10^{-200}$).}
    \label{fig:burst_empirical_results_IET_loglog}
\end{figure}

The aggregate inter-event-time (IET) distribution is heavy-tailed at both levels, with a maximum-likelihood power-law tail~\cite{clauset2009power} of exponent $\alpha \approx 1.8$ that is decisively preferred over an exponential alternative (Figure~\ref{fig:burst_empirical_results_IET_loglog}). Two-parameter alternatives (lognormal, stretched-exponential, truncated power-law) fit marginally better in absolute log-likelihood, as expected from the extra parameter and the finite observation window; we therefore treat $\alpha$ as a tail-scaling summary rather than a claim of strict Pareto behavior. Per-entity medians are consistent with the aggregate fit (median $\alpha \approx 1.9$/$2.1$ and median $B \approx 0.65$/$0.57$ for individuals/groups).

\textbf{Robustness.} Three sensitivity checks qualify the prevalence claim. (i)~Benjamini--Hochberg correction at $q = 0.05$ over the per-entity LR tests lowers the Burst share to $61.2\%$ for individuals and $29.6\%$ for groups (reductions of $5.5$ and $11.1$\,pp); the individual prevalence is essentially preserved, whereas the smaller group sample is more sensitive to the correction. (ii)~Sweeping the burstiness threshold $B \in \{0.2, 0.3, 0.4\}$ at the canonical LR cutoff $p < 0.05$ and $50$-event minimum keeps the analyzable Burst share within $[63.1\%, 68.5\%]$ for individuals and $[34.9\%, 45.1\%]$ for groups, bracketing the canonical $66.8\%/40.7\%$ (a stricter $B \ge 0.5$ lowers them to $56.2\%$ and $26.9\%$); the set of Burst-classified entities is moreover invariant to lowering the minimum-events threshold below $50$, since the canonical rule already requires $\ge 50$ events. (iii)~The finite-size correction of Kim and Jo~\cite{kim2016measuring} shifts the median per-entity $B$ from $0.57$ to $0.68$ for groups (and from $0.65$ to $0.77$ for individuals), i.e., the raw burstiness index is a conservative lower bound on the asymptotic burstiness of these series.

\subsection{Circadian vs. Crisis Modulation (F2)}

We next ask whether either of two natural candidate drivers, namely circadian inactivity and the COVID-19 pandemic, explains a measurable share of the observed burstiness.

\textbf{Circadian inactivity has a near-null effect.} For each valid entity, we recompute $B$ after restricting the event timeline to the active band 07:00--23:00 (matching the simulator's wake window in Section~\ref{subsec:sim-framework}, which covers 98\% of empirical activity), then compare the resulting per-entity burstiness distribution against the unrestricted 24h baseline. The two distributions are nearly indistinguishable (Kullback--Leibler divergence $< 10^{-3}$ at both levels), so we find no evidence that circadian inactivity is a primary contributor to the observed burstiness in this dataset.

\begin{figure}[h]
    \centering
    \includegraphics[width=1.0\linewidth]{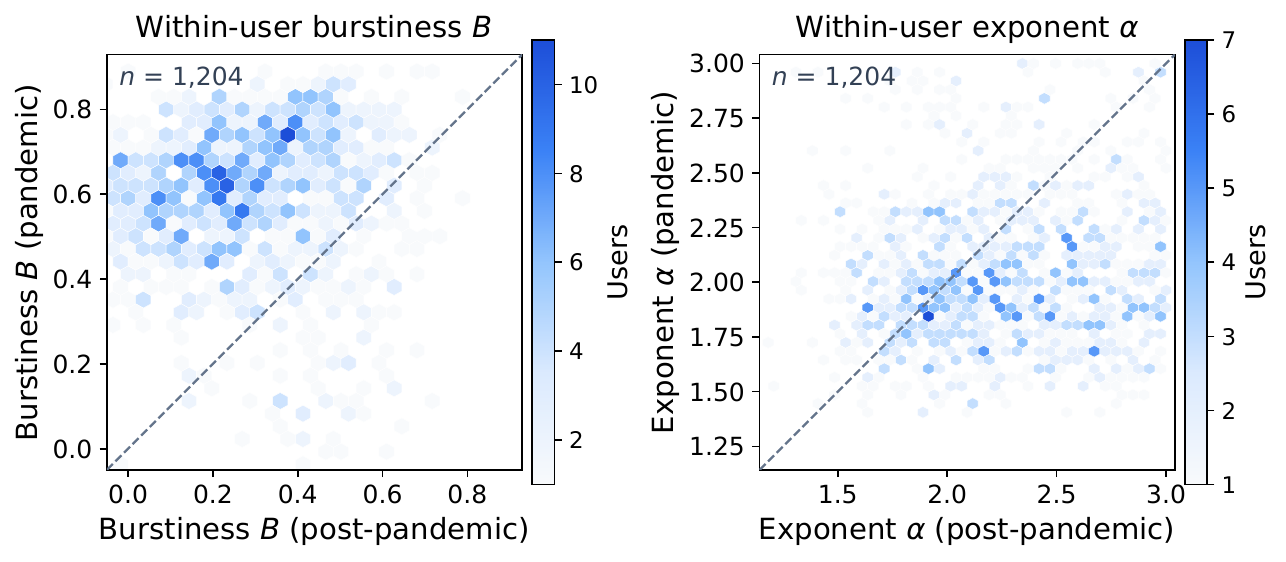}
    \caption{Within-user pandemic vs.\ post-pandemic shift for Burst-classified individuals with $\ge 20$ events in both halves of their own event sequence ($n = 1{,}204$; split at $d = $ 2022-12-08). Each hexbin cell counts users and the dashed line is $y = x$. \textbf{Left:} burstiness index $B$ (pandemic vs.\ post-pandemic); most users lie above the diagonal, i.e.\ are more bursty during the pandemic (median within-user $\Delta B = +0.35$, $86.5\%$ above the diagonal). \textbf{Right:} power-law exponent $\alpha$; most users lie below the diagonal, i.e.\ have a heavier tail (smaller $\alpha$) during the pandemic (median within-user $\Delta\alpha = -0.19$, $66.0\%$ below the diagonal). Both shifts are highly significant under a paired Wilcoxon test ($p < 10^{-130}$ for $B$, $p < 10^{-38}$ for $\alpha$), and a complementary two-sample comparison of the independently re-classified halves agrees in direction and magnitude (Kolmogorov--Smirnov $D = 0.49$, $p < 10^{-44}$ for $B$; $D = 0.33$, $p < 10^{-19}$ for $\alpha$).}
    \label{fig:burst_empirical_results_pandemic_effect}
\end{figure}

\textbf{The pandemic shock yields a substantial shift.} To test the pandemic effect while controlling for cohort composition, we use a \emph{within-user paired} design: for each Burst individual with at least $20$ events in both the pandemic half ($P$, $t < d$) and the post-pandemic half ($Q$, $t \ge d$) of its own sequence (split at the reopening date $d=\text{2022-12-08}$, Section~\ref{subsec:data}), we recompute $B$ and $\alpha$ on each half and pair the two measurements within the user ($n = 1{,}204$ users). Pairing each user against itself removes the user-base churn that confounds a between-cohort comparison, which is a real concern here because the $\sim 10$-month post-pandemic window is far shorter than the $\sim 33$-month pandemic window. Both burstiness measures shift substantially and highly significantly toward the pandemic half: burstiness rises (median within-user $\Delta B = +0.35$) and the tail grows heavier ($\Delta\alpha = -0.19$), and a complementary two-sample comparison of the independently re-classified halves agrees in direction and magnitude (Figure~\ref{fig:burst_empirical_results_pandemic_effect}). At the group level the direction is the same but the burstiness shift is not statistically significant ($B$ two-sample KS $p = 0.18$), and the paired design is infeasible because only $3$ groups remain Burst-classified in both halves (out of $1{,}228$ pandemic-Burst groups); the group-level comparison is thus severely underpowered by the same window-length asymmetry. We therefore treat the within-user individual result as the primary evidence for the pandemic effect.

\subsection{Endogenous Self-Excitation and Memory Structure (F3)}
\label{sec:empirical_mechanism}

Having established that the burstiness is real and crisis-amplified, we ask what generates it. Two complementary diagnostics point to endogenous self-excitation.

\begin{figure}[h]
    \centering
    \includegraphics[width=0.8\linewidth]{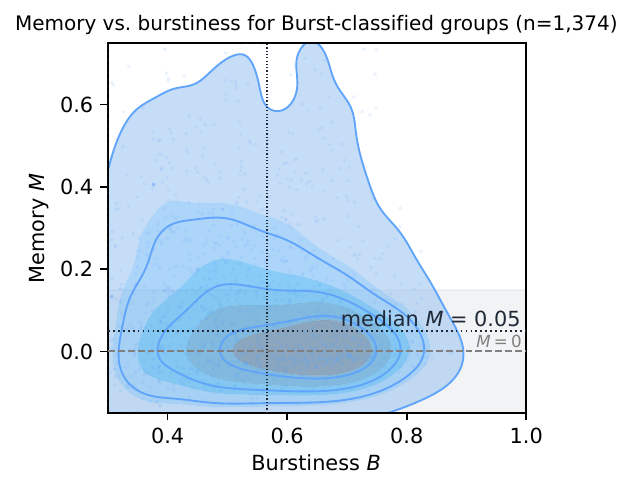}
    \caption{Memory vs.\ burstiness for Burst-classified group trajectories ($n = 1{,}374$). Each point is one tracked group; the 2D kernel-density estimate shows concentration at high $B$ and $M \approx 0$ (near-memoryless lag-1 inter-event-time correlation). Dashed lines mark $M = 0$, the burstiness threshold $B = 0.3$, and the cohort median $B$.}
    \label{fig:burst_empirical_results_BM_plane}
\end{figure}

\textbf{The data sit in the near-memoryless burstiness region.} Following Goh and Barab\'{a}si~\cite{goh2008burstiness}, we characterize each Burst-classified group by both the burstiness index $B$ and the memory coefficient $M$, defined as the lag-1 Pearson correlation of consecutive inter-event times. Across the $1{,}374$ Burst-classified group trajectories (entities with $\ge 50$ events), the median $(B, M) = (0.57, 0.05)$ and $\approx 69\%$ fall within $|M| \le 0.15$ (Figure~\ref{fig:burst_empirical_results_BM_plane}). The data therefore concentrate in the $(+B, M \approx 0)$ region of the plane, the same region occupied by online communication and web-browsing activity~\cite{goh2008burstiness, barabasi2005origin}, and clearly separated from regimes with strongly positive $M$ (geophysical signals, library borrowings) or strongly negative $M$ (priority-queueing-dominated tasks). Operationally, this places the participation timing in a regime where heavy-tailed dwell times coexist with near-memorylessness between consecutive intervals, a regime well described by self-exciting Hawkes-type cascades whose individual cascades terminate quickly relative to the long inactive intervals between them.

\textbf{Most bursty entities are self-exciting.} Applying the Hawkes self-excitation test of Section~\ref{subsec:burst-quantification} (\emph{self-excited} if $p_\text{LRT} < 0.05$ and $\hat\alpha > 0$) and restricting to the Burst-classified records, $79.7\%$ of individuals and $69.3\%$ of groups qualify as self-exciting. This confirms that the heavy-tailed timing is driven predominantly by endogenous self-excitation rather than Poisson-like clustering, and, together with the dissociation in F2, identifies the two mechanisms the simulator must reproduce: a self-excitation backbone and a crisis-period amplification. The per-user Hawkes parameters that quantify these mechanisms are the calibration source for the simulator (Section~\ref{subsec:hawkes-fit}).

\section{A When/What-Decoupled Simulator for Bursty Participation}
\label{sec:methods}

We now turn from data to simulation. We first describe the city-scale environment and the synchronous LLM baseline (Section~\ref{subsec:sim-framework}). We then introduce a dual-channel decision protocol that gates the LLM with the empirically-fitted self-excitation mechanism of F3 (Section~\ref{subsec:sim-modes}), calibrate it to the bounded simulated horizon (Section~\ref{subsec:hawkes-fit}), and test whether it is sufficient to lift per-agent timing above the burst threshold (Section~\ref{subsec:sim-results}).

\subsection{Simulation Environment and the Synchronous Baseline}
\label{subsec:sim-framework}

\begin{figure}[h]
    \centering
    \includegraphics[width=0.95\linewidth]{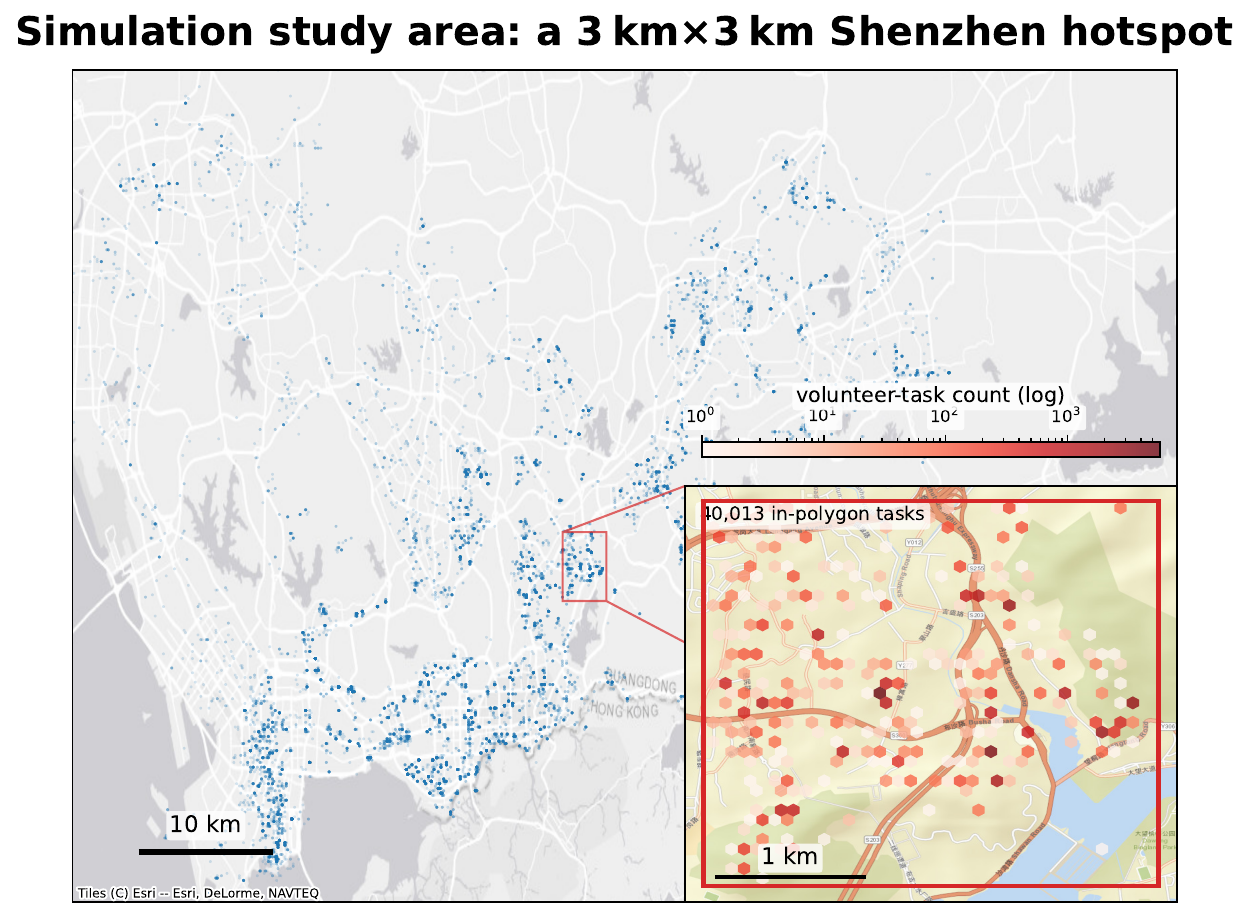}
    \caption{Simulation study area. All agents are confined to a $3\,\text{km}\times3\,\text{km}$ hotspot polygon (red box) centered on the densest cell of the Shenzhen volunteer-task pool; blue dots show the city-wide distribution of volunteer tasks. The bottom-right inset zooms into the hotspot and overlays the spatial density (log scale) of the $40{,}013$ in-polygon tasks used to build the simulation task stream. }
    \label{fig:burst_study_area}
\end{figure}

\paragraph{Environment and agents.}
We build a city-scale volunteer simulation on AgentSociety~\cite{piao2025agentsociety}, which provides an LLM-driven agent runtime, a real-map environment with point-of-interest (POI) and area-of-interest (AOI) annotations, and a step-wise workflow engine. To match the spatial scale of empirical clustering, we restrict all agents to a $3\,\text{km} \times 3\,\text{km}$ hotspot polygon centered on the densest task cell of the Shenzhen task pool (Figure~\ref{fig:burst_study_area}), which contains 122 in-polygon AOIs. We instantiate $N_\text{agent} = 50$ \emph{passive} citizen agents that take no autonomous per-step action and act only when queried by the simulator, drawn from the AgentSociety default citizen profile pool; mobility and social blocks are enabled while economy-related blocks are disabled so the simulation focuses on volunteering. All agents share a single LLM backend (\textit{GPT-5-mini}).

\paragraph{Task pool, time, and decision schedule.}
The task environment is a hotspot-filtered subset of the empirical task table, preserving each task's start/end time, location, type, and capacity. For each simulated day we draw $K_\text{day} = \lfloor 0.35 \cdot N_\text{agent} \rceil = 18$ tasks, where the ratio $0.35$ is the empirical median of (daily unique tasks)/(daily unique active volunteers) within the hotspot polygon. Time is discretized into hourly steps (one step equals one hour). We define quiet hours as 23:00--07:00, during which no participation decisions are queried, mirroring the empirical concentration of volunteer tasks in daytime hours (the active band 07:00--23:00 covers $98\%$ of empirical activity). At each non-quiet hour, every agent receives the set of tasks whose active interval overlaps a two-hour look-ahead window and for which it is \emph{eligible}: its current busy state must not exceed the task start and the task's remaining capacity must be positive.

\paragraph{Prompt and action format.}
In every mode the LLM is asked to return a JSON object \texttt{\{"participate": bool, "task\_index": int\}} given the current day and hour and a numbered list of eligible tasks (each annotated with type, status, location, time window, capacity, and current participant count), and is forbidden from participating in more than one task at a time. The Baseline mode uses a conservative system prompt that instructs the agent to decline when uncertain. The two Hawkes-enabled modes instead use a gate-aligned system prompt that removes this blanket ``decline-if-uncertain'' default and asks the agent to confirm sign-up whenever at least one listed task is reasonable, and additionally append a short \emph{qualitative} participation-context sentence derived from the agent's own recent accepted-event history (how recently it last participated, whether it is in an active streak) and from whether the current step falls in a crisis-boost window. Crucially, no numeric Hawkes quantity ($\mu$, $\alpha$, $\beta$, or the intensity $\lambda^\text{eff}$) is ever exposed to the LLM: the temporal mechanism enters the LLM channel only through this qualitative recency wording and through the gating mechanism of Section~\ref{subsec:sim-modes}, never as an explicit rate.

\paragraph{Implementation details.}
Each agent's home and work AOI are randomly remapped to in-polygon AOIs at simulation start. The per-day task draw uses reservoir sampling with the random seed offset by the day index for reproducibility. Eligibility is tracked through a per-agent \texttt{busy\_until} timestamp, checked against each task's own active window; whenever a participation is accepted the simulator schedules a walking trip from the agent's current AOI to the in-polygon AOI nearest the task and a return walk after the task ends.

\paragraph{Why a synchronous LLM simulator is not bursty.}
The Baseline simulator is representative of standard LLM-based multi-agent practice: at every non-quiet hour each eligible agent is prompted independently, and its decision depends only on the currently listed tasks, not on \emph{when} it last acted. This per-agent process is close to independent hourly Bernoulli trials with no channel by which one participation raises the probability of the next, so it has neither the endogenous self-excitation (F3) nor the crisis-period amplification (F2) found in the data. We therefore expect it to fail the burst criterion of Eq.~\eqref{eq:burst-rule} and use it as the negative-control baseline against which the mechanism-augmented protocol below is measured (confirmed in Section~\ref{subsec:sim-results}).

\subsection{Mechanism-Augmented Decision Protocol}
\label{subsec:sim-modes}

\begin{figure}[h]
    \centering
    \includegraphics[width=0.95\linewidth]{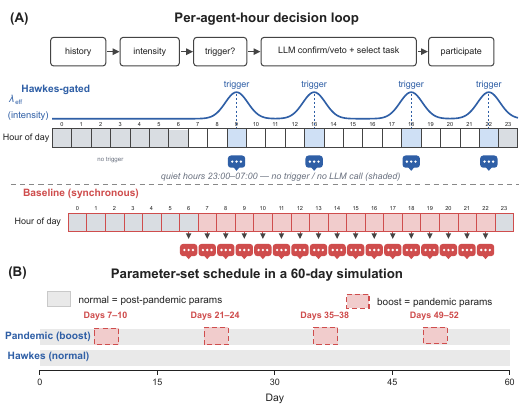}
    \caption{Simulation mechanism. \textbf{(A)} Per-agent-hour decision loop over a day. In the \emph{Baseline (synchronous)} mode, every eligible agent is prompted at each non-quiet hour and the LLM alone decides participation. In the \emph{Hawkes-gated} modes, the agent's self-exciting intensity $\lambda^\text{eff}_u(t)$, built from its accepted-event history, gates each hour: the hour \emph{triggers} only when the thinning sampler draws at least one event, and only a triggered hour issues an LLM call to confirm/veto and select a task before the agent participates. Quiet hours (shaded, 23:00--07:00) and non-triggered hours issue no LLM call. \textbf{(B)} Parameter-set schedule of the two Hawkes-gated modes over the 60-day horizon. \emph{Hawkes (normal)} uses the normal (post-pandemic-window) set throughout, whereas \emph{Pandemic (boost)} inserts the boost (pandemic-window) set on days 7--10, 21--24, 35--38, and 49--52 over the same normal backbone.}
    \label{fig:sim_mechanism}
\end{figure}

We compare three participation modes that share an identical environment, task stream, agent pool, and per-day task sampling; they differ only in whether a temporal-mechanism channel is present and, when present, in which Hawkes parameter set it uses. The two channels below are combined through a \emph{gate} coupling: the Hawkes process supplies the timing, and the LLM confirms (or vetoes) and selects the task.

\paragraph{Channel A: LLM decision.}
We denote the LLM's binary participation indicator by $a^\text{LLM}_{u,t} \in \{0, 1\}$ and its task choice by $k^\text{LLM}_{u,t} \in \{1, \dots, K^{e}_{u,t}\}$, where $K^{e}_{u,t}$ is the number of eligible tasks for agent $u$ at hour $t$. In Baseline, an LLM call is issued at every non-quiet hour for every agent with at least one eligible task. In the Hawkes-enabled modes the LLM is queried \emph{only} on agent-hours for which Channel~B fires (see below), so the LLM serves as a confirm/veto and task selector for events whose timing has already been proposed by the Hawkes channel.

\paragraph{Channel B: Hawkes timing gate.}
For Hawkes-enabled modes, each agent-hour is gated by its per-agent intensity $\lambda^\text{eff}_u(t)$, computed from the agent's Hawkes parameters $(\mu_u, \alpha_u, \beta_u)$ and accepted-event history $H_u(t) = \{t^u_j \le t\}$ via~\eqref{eq:hawkes_individual}. Rather than a single hourly Bernoulli draw, we sample the second-resolution event times the intensity produces within the hour by Ogata thinning, yielding a (possibly empty) minute-scale cluster; the agent-hour \emph{triggers} iff at least one event is sampled, which occurs with probability
\begin{equation}
\Pr\!\bigl(\text{trigger}_{u,t}\bigr) = 1 - \exp\!\bigl(-\Lambda^\text{eff}_u(t)\bigr),\qquad
\Lambda^\text{eff}_u(t) = \!\!\int_{t}^{t+1\text{h}}\!\!\lambda^\text{eff}_u(s)\,ds .
\label{eq:hawkes_trigger}
\end{equation}
This sub-hour sampling produces short intra-cluster inter-event times that, together with the long between-cluster gaps, elevate the burstiness index and preserve the heavy inter-event-time tail, the timing signature the burst criterion is designed to detect.

\paragraph{Aggregation (gate coupling).}
As summarized in Figure~\ref{fig:sim_mechanism}A, a non-triggered agent-hour is skipped (no LLM call), whereas a triggered hour issues a single LLM call that confirms or vetoes the sampled minute-scale cluster; the agent therefore participates iff the Hawkes channel triggers \emph{and} the LLM confirms:
\begin{equation}
\Pr(\text{participate}_{u,t}) = \bigl(1 - e^{-\Lambda^\text{eff}_u(t)}\bigr)\; p^\text{LLM}_{u,t},
\label{eq:dual_channel}
\end{equation}
where $p^\text{LLM}_{u,t}$ is the LLM confirmation probability under the gate-aligned prompt. The selected task index is $k^\text{LLM}_{u,t}$ when the LLM confirms and the index is feasible; if the LLM confirms but returns an out-of-range index, we choose uniformly among the agent's eligible tasks. Routing the temporal mechanism through the gate (rather than overriding the LLM) keeps the LLM's content-based judgement in the loop while letting the Hawkes process control \emph{when} decisions are made, and collapses the LLM call budget to the number of triggered agent-hours.

\paragraph{Acceptance feedback and feasibility.}
Whenever a participation is accepted, the task start time is appended to $H_u$, \texttt{busy\_until} is updated to the task end, and the task's remaining capacity is decremented, so the self-excitation in~\eqref{eq:hawkes_individual} compounds over the agent's own accepted history. If the gated participation cannot be placed on any eligible task (e.g.\ all candidates are full or conflict with \texttt{busy\_until}), the event is dropped.

\paragraph{Three modes.}
We use a dual parameter-set switching design: each agent carries two empirically-grounded Hawkes parameter sets, a \emph{normal} set derived from the post-pandemic window and a \emph{boost} set derived from the pandemic window (Section~\ref{subsec:hawkes-fit}).
\begin{itemize}
\item \textbf{Baseline (LLM-only).} Channel B is disabled; the LLM is queried at every non-quiet active hour and the participation decision is the LLM output alone.
\item \textbf{Hawkes (normal).} Channel B is enabled with gate coupling and the agent always uses its \emph{normal} (post-pandemic) parameter set: $\lambda^\text{eff}_u(t) = \lambda_u(t \mid \mu_u^{Q}, \alpha_u^{Q}, \beta_u^{Q})$ throughout.
\item \textbf{Pandemic (boost).} This windowed-boost mode alternates between its \emph{boost} (pandemic) and \emph{normal} (post-pandemic) parameter sets on a periodic schedule (Figure~\ref{fig:sim_mechanism}B): inside the boost windows $\mathcal{T}_\text{boost}$ it uses $(\mu_u^{P}, \alpha_u^{P}, \beta_u^{P})$ and outside it reverts to $(\mu_u^{Q}, \alpha_u^{Q}, \beta_u^{Q})$, applying a 4-day crisis burst within each 14-day cycle so that the crisis regime is active well under half of the 60-day horizon. The boost set elevates both the baseline intensity and the self-excitation strength together (Section~\ref{subsec:hawkes-fit}), so the crisis perturbation acts jointly on activation and triggering rather than on the activation rate alone, preserving the empirically observed coupling between elevated baseline activity and stronger triggering during the crisis window.
\end{itemize}

\subsection{Horizon-Aware Hawkes Calibration}
\label{subsec:hawkes-fit}

The gate of Section~\ref{subsec:sim-modes} is parameterized entirely from the data. To supply the \emph{boost} and \emph{normal} regimes we fit the univariate Hawkes process~\eqref{eq:hawkes_individual} per user, separately on the pandemic window $P$ and the post-pandemic window $Q$ (Section~\ref{subsec:data}). Fitting on the windows rather than the full horizon is necessary: because the full 3.6-year horizon spans pandemic onset, prolonged restrictions, and a post-restriction normalization, the stationarity assumption underlying~\eqref{eq:hawkes_individual} fails when applied to the entire log; pilot fits on the full horizon yield median branching ratio $\hat\eta\!=\!\hat\alpha_u/\hat\beta_u\!\approx\!0.92$, indicative of structural non-stationarity absorbed into the kernel. Within each window we require at least 20 events and obtain $(\hat\mu_u, \hat\alpha_u, \hat\beta_u)$ by joint maximum likelihood: a grid over $(\eta, \beta)$ with $\eta \in [0.02, 0.98]$ and $\beta$ logarithmically spaced around the user's median inter-event time, followed by localized random search, with adequacy checked by the time-rescaling KS diagnostic of Section~\ref{subsec:burst-quantification}. The per-user pandemic- and post-pandemic-window fits, each split into a self-exciting subset (the bursty pool) and a Poisson subset, form the simulation parameter pools used below.

\paragraph{Per-agent parameters and horizon-aware calibration.}
Each simulated agent is assigned, at simulation start, a paired bundle drawn from the per-user pandemic- and post-pandemic-window Hawkes fit tables, intersected on user identity. A user is placed in the \emph{bursty pool} if its pandemic-window fit is self-exciting ($p_\text{LRT} < 0.05$) and in the \emph{Poisson pool} otherwise; a fraction $r = 0.66$ of agents receive bursty bundles, matching the empirical individual Burst share of $66.8\%$ (Section~\ref{sec:empirical_findings}), and the rest receive Poisson bundles. Because the simulated horizon (tens of days) is orders of magnitude shorter than the multi-year empirical window, the empirical \emph{absolute} event rates cannot be transferred verbatim without either saturating the timeline or starving agents below the classification threshold. We therefore re-parameterize each bundle to the simulated horizon while preserving the empirical self-excitation \emph{shape}: for bursty agents the branching ratio $\eta_u = \alpha_u/\beta_u$ is inherited from the fit and clipped to a near-critical band $[0.85, 0.97]$; we then draw a per-agent target event count from $[40,80]$ (bursty) or $[15,30]$ (Poisson), set the kernel timescale from the resulting event rate $\Lambda = \text{target}/T_\text{active}$ via $\beta_u = \kappa\,\Lambda$ with separation ratio $\kappa = 10$ and $\alpha_u = \eta_u \beta_u$, and solve the baseline rate $\mu_u$ in closed form from $\mathbb{E}[N] = \mu_u T_\text{active}/(1-\eta_u)$ so the expected volume matches the target. The boost regime of a bursty agent uses $1.5\times$ this target (scaling $\mu_u, \alpha_u, \beta_u$ together while leaving $\eta_u$ and the separation ratio invariant), so crisis windows are both hotter and burstier. Choosing $\beta_u$ from the rate makes the excitation memory $1/\beta_u \approx (1/\kappa)$ of the mean inter-event gap, keeping clusters temporally distinct (observable $B > 0.3$) instead of smearing into a near-constant, near-Poisson intensity. To remove the cold-start gap that would otherwise pin early-horizon medians below the minimum-events threshold, each bursty agent is seeded with $2$ warm-start events placed just before the first active hour of day~1.

\begin{table*}[h]
    \centering
    \small
    \caption{Simulation results for the three participation modes and the LLM ablation ($N_\text{agent}=50$, 60-day horizon). \textbf{LLM} \emph{on}/\emph{off} marks the gate-only ablation: when \emph{off}, every Hawkes-triggered agent-hour auto-participates with the LLM bypassed (Baseline has no Hawkes gate). \emph{Qualifying} agents have $\geq\!20$ task-level events (relaxed from the empirical threshold of $50$ given the bounded horizon); \emph{Bursty (\%)}, \emph{Median $B$}, and \emph{Median $\alpha$} are computed over them under the joint criterion ($B>0.3$, LR $p<0.05$), and \emph{Mean events/agent} is averaged over all $50$ agents. The per-agent $B$ distribution of each gate-only arm is statistically indistinguishable from its LLM-coupled counterpart ($B$ Mann--Whitney $p=0.70$/$0.57$ for Hawkes/Pandemic); the pooled IET distributions are likewise similar for Pandemic (boost) (KS $p=0.15$) and only marginally different for Hawkes (normal) (KS $p=0.03$), so the burstiness is governed by the calibrated Hawkes gate rather than the LLM. }
    \label{tab:sim_results}
    \setlength{\tabcolsep}{8pt}
    \renewcommand{\arraystretch}{1.2}
    \begin{tabular}{llrrrrr}
    \toprule
    \textbf{Mode} & \textbf{LLM}
      & \textbf{Qualifying agents}
      & \textbf{Mean events/agent}
      & \textbf{Bursty (\%)}
      & \textbf{Median $B$}
      & \textbf{Median $\alpha$} \\
    \midrule
    Baseline (no gate) & on  & 45 & 62.0 & 0.0  & \mbox{$-0.14$} & n/a \\
    \midrule
    Hawkes (normal)    & on  & 8  & 13.5 & 50.0 & 0.37 & 1.45 \\
    Hawkes (normal)    & off & 13 & 16.0 & 61.5 & 0.39 & 1.40 \\
    \midrule
    Pandemic (boost)   & on  & 15 & 16.8 & 66.7 & 0.37 & 1.36 \\
    Pandemic (boost)   & off & 16 & 17.7 & 56.2 & 0.38 & 1.36 \\
    \bottomrule
    \end{tabular}
    \end{table*}

\paragraph{Logging.}
For every queried agent-hour we log the parsed decision, raw model output, and participation mode; Hawkes-enabled modes additionally record $\lambda^\text{eff}_u(t)$, the per-hour trigger probability and outcome, the number of sub-hour events sampled, and the gate decision source (LLM-confirmed, random-task fallback, or LLM veto). These records, with the sampled daily tasks and hourly schedule, feed the temporal and group-level analyses that follow.

\subsection{Evaluation}
\label{subsec:sim-results}

We deploy $50$ citizen agents in the Shenzhen hotspot environment over a 60-day horizon with an hourly decision interval; participation decisions are queried only during active hours (no decisions during quiet hours 23:00--07:00), each agent participates in at most one task at a time, and participation is limited by task capacity. Burstiness is evaluated with the same task-event-level criterion used for the empirical data (Section~\ref{subsec:burst-quantification}), with the minimum-events threshold relaxed from $50$ to $20$ because the bounded simulated horizon caps per-agent event volume. Table~\ref{tab:sim_results} summarizes the results.

\paragraph{F4 (necessity): a synchronous LLM simulator is structurally non-bursty.}
The Baseline (LLM-only) mode produces \emph{no} bursty agents (Table~\ref{tab:sim_results}): its per-agent timing is near-regular (median $B<0$), exactly as predicted in Section~\ref{subsec:sim-framework}. This deficit is not for lack of activity---Baseline has by far the highest mean event count---nor for lack of compute: querying the LLM at every active hour, it spends an order of magnitude more LLM calls than the gated Hawkes channels ($\sim$28k vs.\ $\sim$0.7--0.9k, a $\approx 30\times$ gap, since the gated channels invoke the LLM only on triggered agent-hours) yet still collapses to sub-Poisson timing. Temporal realism is thus governed by \emph{when} decisions are made, not by how many LLM calls or events a mode generates.

\paragraph{F5 (feasibility): a calibrated gate is sufficient to cross the burst threshold.}
Adding the gated Hawkes channel lifts the bursty share from the Baseline's $0\%$ to $50.0\%$ for Hawkes (normal) and $66.7\%$ for Pandemic (boost), with median $B$ comfortably above the threshold in both modes (Table~\ref{tab:sim_results}). The crisis regime additionally raises the number of classifiable agents and the mean event count, consistent with the elevated baseline intensity of the pandemic-window parameter set. The two Hawkes modes differ mainly in bursty-agent share rather than within-agent $B$: the boost rescales the event rate while leaving the self-excitation shape (branching ratio and cluster separation) unchanged, and $B$ is invariant to such rescaling (Section~\ref{subsec:sim-modes}). The roughly flat event-count column further confirms that this gain is not an artefact of differential event-count eligibility across modes.

\paragraph{F6 (separability): timing is set by the gate, content by the LLM.}
The Baseline-vs-Hawkes contrast shows that the gate is \emph{necessary} for burstiness, but it does not by itself rule out an interaction in which the LLM shapes the timing at gate-triggered moments. By construction the LLM can only confirm or veto an entire Hawkes-proposed cluster and never alters the intra-cluster spacing, so this test mainly probes whether veto decisions reshape the between-cluster pattern. To isolate the source of the simulated burstiness we re-ran both Hawkes modes with the LLM channel bypassed (the \emph{gate-only} ablation: every Hawkes-triggered agent-hour auto-participates on a randomly chosen eligible task), holding the sample seed and every Hawkes and calibration knob fixed. The gate-only arms reproduce the LLM-coupled arms' burstiness: the per-agent $B$ distributions are statistically indistinguishable and the pooled inter-event-time distributions are close, with comparable bursty shares and median $B$ (test statistics in Table~\ref{tab:sim_results}). Removing the LLM thus neither flattens nor sharpens the per-agent burstiness, consistent with the temporal structure being governed by the empirically-calibrated Hawkes gate. The LLM's contribution is consequently \emph{orthogonal} to timing: it acts as a context-dependent veto (realized events per agent are slightly \emph{lower} in the LLM-coupled arms) and selects \emph{which} task and area a triggered agent joins, neither of which a bare Hawkes sampler can supply. 

\section{Discussion}
This paper argues that a realistic LLM-based social simulator must capture not only plausible individual decisions but also realistic temporal dynamics, a dimension that standard synchronous simulators miss. A synchronous LLM baseline, which lacks any self-excitation channel, produces reasonable actions yet collapses to near-regular timing, and issuing far more LLM calls does not create the missing mechanism. Routing the temporal decision through an empirically-fitted Hawkes gate, together with a separate crisis-period regime, supplies that mechanism and lifts per-agent timing above the burst threshold. A gate-only ablation (Section~\ref{subsec:sim-results}) leaves per-agent burstiness essentially unchanged when the LLM is bypassed, confirming that the timing is fixed by the gate rather than the LLM. The two are therefore separable by construction: the gate governs \emph{when} agents act, while the LLM decides \emph{what} they do, namely which task a triggered agent joins and whether a capacity-constrained opportunity is worth taking.

The empirical analysis also revises a common assumption in the bursty-dynamics literature. Of the two canonical candidate drivers we tested, a daytime-only restriction leaves the burstiness distribution essentially unchanged, whereas a within-user pandemic vs.\ post-pandemic comparison reveals a substantial and direction-consistent elevation of individual burstiness. This dissociation, together with the position of the data in the near-memoryless region of the Goh--Barab\'{a}si plane, suggests that the heavy-tailed timing in offline volunteering is driven primarily by exogenous crisis-period activation acting on a near-memoryless self-excitation backbone, rather than by passive circadian inactivity. It is this dissociation that motivates our simulation design, which combines a Hawkes channel for the self-excitation backbone with a pandemic-window parameter set for the crisis perturbation, instead of relying on an alternative such as a hard-coded day/night activity schedule.

The practical value of this framework lies in its ability to support low-cost experimentation on crisis mobilization mechanisms. For social scientists, it provides a controllable environment for testing how external shocks, priority changes, and endogenous reinforcement may shape collective participation. For platform designers and policy makers, such simulations can help evaluate potential interventions before they are deployed in real communities, such as prioritizing urgent tasks, adjusting notification strategies, or anticipating sudden increases in volunteer supply. In this sense, LLM-based social simulation can become a complementary tool for studying human collective behavior when field experiments are costly, slow, or ethically difficult.

\section{Conclusion}
In this study, we characterized bursty participation in a city-scale Shenzhen volunteering log at both the individual and tracked-group levels, showing that heavy-tailed timing is prevalent and robust, that circadian inactivity barely affects it while the pandemic period substantially elevates individual burstiness, and that most bursty entities are endogenously self-exciting. Guided by these findings, we built a mechanism-augmented LLM city simulator that decouples \emph{when} an agent acts, via an empirically-fitted Hawkes gate and a pandemic-window regime, from \emph{what} it does, decided by the LLM. As a proof-of-concept, this gate is sufficient to lift per-agent timing across a burst threshold that an LLM-only baseline never reaches. Because our simulation is limited to a single hotspot of 50 agents over a bounded horizon, scaling the mechanism to larger populations and validating non-calibrated targets, such as the tail exponent and group-level burstiness, are natural next steps.


\bibliographystyle{IEEEtran}
\bibliography{IEEEabrv,IEEEexample}

\end{document}